\documentclass[aps,prl,twocolumn,showpacs,superscriptaddress,floatfix]{revtex4}
\usepackage[dvips]{graphicx}
\usepackage[english]{babel}
\usepackage{color,enumerate,amsmath,amssymb,amsbsy,amscd,bm}
\usepackage{pstricks,epsf,epsfig}
\usepackage{epstopdf}

\newcommand{\ket}[1]{|{#1} \rangle}  
\newcommand{\bra}[1]{\langle {#1}|}  
\newcommand{\ketbra}[2]{|{#1}\rangle\langle{#2}|} 
\newcommand{\bs}{{\mathbf s}}   
\newcommand{\bS}{{\mathbf S}}   

\begin{document}
\preprint{}
\title[]{Heisenberg Spin Bus as a Robust Transmission Line for Perfect State Transfer}
\author{Sangchul Oh}
\affiliation{Department of Physics, University at Buffalo,
State University of New York, Buffalo, New York 14260-1500, USA}
\author{Lian-Ao Wu}
\affiliation{Department of Theoretical Physics and History of Science,
The Basque Country University (EHU/UPV), PO Box 644, 48080 Bilbao, Spain}
\affiliation{IKERBASQUE, Basque Foundation for Science, 48011 Bilbao, Spain}
\author{Yun-Pil Shim}
\affiliation{Department of Physics, University of Wisconsin-Madison,
Madison, Wisconsin 53706, USA}
\author{Mark Friesen}
\affiliation{Department of Physics, University of Wisconsin-Madison,
Madison, Wisconsin 53706, USA}
\author{Xuedong Hu}
\affiliation{Department of Physics, University at Buffalo,
State University of New York, Buffalo, New York 14260-1500, USA}
\date{\today}
\begin{abstract}
We study the protocol known as quantum state transfer for a strongly coupled 
antiferromagnetic spin chain or ring (acting as a spin bus), with weakly coupled 
external qubits. By treating the weak coupling as a perturbation, we find that 
\emph{perfect} state transfer (PST) is possible when second order terms are 
included in the expansion.  We also show that PST is robust against variations in 
the couplings along the spin bus and between the bus and the qubits. As evidence 
of the quantum interference which mediates PST, we show that the optimal time for 
PST can be \emph{smaller} with larger qubit separations, for an even-size chain 
or ring.
\end{abstract}
\maketitle

Transferring quantum states between qubits is one of the most fundamental tasks 
in quantum information processing.  State transfer can be realized through various 
approaches. For example, if two qubits have a controllable Heisenberg coupling, 
a swap gate can be used to exchange their quantum states~\cite{Loss98}. For longer 
range communications, a quantum data bus connecting remote qubits can be employed: 
examples include phonon modes for trapped ions~\cite{Cirac95}, cavity photon modes 
for superconducting qubits~\cite{Blais04}, and spin chains for spin 
qubits~\cite{Bose03,Friesen07}.

The feasibility of quantum state transfer through spin chains or lattices depends 
on the available types of coupling and the degree of controllability. One of the 
most widely studied schemes is known as perfect state transfer (PST). In this 
protocol, the bus and qubits are prepared in an initial state; the system is then 
allowed to evolve freely, with all the spin interactions held constant. State transfer 
cannot be achieved with perfect fidelity in a uniform spin chain with Heisenberg or 
XY spin couplings~\cite{Bose03}, except for individual couplings that are specially 
engineered and non-uniform~\cite{Christandl04, Wu09, Kay_PRL10, Lukin10, Bose11}. 
The quantum information ordinarily propagates dispersively, with excited states of 
the system playing a critical role in the evolution. This is in contrast with 
the adiabatic operating modes of the spin bus, particularly the odd-size 
bus~\cite{Friesen07}, and it leads to fundamental questions about how the excited
states of a spin lattice carry quantum information: whether the propagation of quantum 
information is limited by the speed of the spin wave~\cite{Eisert09}, and whether the 
quantum state transfer time is simply proportional to the distance between sender and 
receiver, in analogy with signal transmission through wires or optical fibers.

In this paper, we return to the question of whether it is possible to implement PST 
between remote qubits through spin-1/2 Heisenberg chains or rings with uniform couplings.
In this geometry, we show that if the qubit-chain coupling is smaller than the 
intra-chain coupling (i.e. a spin bus architecture~\cite{Friesen07}), PST can be 
achieved with arbitrary accuracy. Further, we show that, despite significant differences 
in the ground state properties of odd and even-size chains or rings, both systems admit 
similar types of PST, with fidelities that are insensitive to variations in the individual 
exchange couplings in the system. These results suggest that we can view a uniformly 
coupled Heisenberg chain as a robust quantum coherent plug-in device between two spin 
qubits, like a transmission line between two circuit elements, so that arbitrary quantum
states may be swapped after a fixed time determined by the coupling parameters. 
In addition, we show that quantum interference effects are inherent in the state transfer 
protocol by demonstrating that the transfer time is not simply proportional to the 
geometric distance between the source and target qubits. This counterintuitive observation 
indicates that quantum information transfer through a quantum spin transmission line
cannot be fully understood in terms of a classical analog, such as a signal propagating 
along a metal wire.

\begin{figure}[htbp]
\includegraphics[scale=0.5]{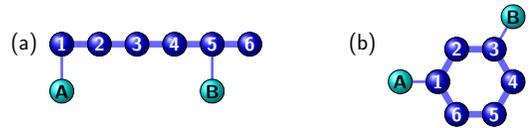}
\caption{(Color online). PST in the spin bus geometry, where two qubits are coupled to 
(a) an even or odd spin-chain, or (b) an even size ring.  The qubit-bus couplings are 
taken to be weak compared to intra-bus couplings.}
\label{Fig1}
\end{figure}

The system we consider consists of two qubits $A$ and $B$ weakly attached to a strongly 
coupled Heisenberg chain or ring, as shown in Fig.~{\ref{Fig1}}. The total Hamiltonian 
is
\begin{align}
H = H_C + H_{QC}\,,
\label{Hamil}
\end{align}
where the Heisenberg Hamiltonian of the chain or ring is
\begin{align}
H_C & = \sum_{i=1}^{N} J_i\, \bs_i\cdot\bs_{i+1}\,.
\label{Hamil_chain}
\end{align}
We first assume that the chain or ring has uniform antiferromagnetic couplings, with 
$J_i= J_0 > 0$ for $i=1,\cdots, N-1$. The boundary conditions are $J_N = 0$ for an open 
chain, or $J_N=J_0$ and $\bs_{N+1} \equiv \bs_{1}$ for a ring. The couplings between 
qubits $A$ and $B$ and the $i$-th and $j$-th spins of the chain are given by
\begin{align}
H_{QC} = J_{A,i}\, \bS_A\cdot\bs_{i} + J_{B,j}\, \bS_B \cdot\bs_{j}\,.
\label{Hamil_coupling}
\end{align}
The qubit-chain coupling is assumed to be weak, with 
$0< \lambda_{\alpha} \equiv J_{\alpha,i}/J_0 \ll 1$.  Here, $\alpha =A, B$, and
$\lambda_{\alpha}$ will be used as a perturbation parameter. From now on, we set 
$J_0 = \hbar = 1$.  Note that we have previously studied the low-energy effective 
Hamiltonian arising from Eq.~(\ref{Hamil})~\cite{Oh10}. Here, we focus on PST between 
qubits $A$ and $B$, by numerically solving its time evolution.

Let us first define the state transfer protocol for the spin bus geometry. The goal 
is to transfer an unknown quantum state from qubit $A$ to qubit $B$ through chain 
$C$ in a fixed time period, during which the spin couplings are held constant. 
At $t=0$, an initial state of the total system is prepared as 
$\ket{\Psi(0)} 
= \bigl(\, a\ket{0_A} + b\ket{1_A} \,\bigr) \otimes \ket{0_C} \otimes \ket{0_B}$, 
where qubit $A$ is in a superposed pure state represented by a point on a Bloch 
sphere with $a = \cos\frac{\theta}{2}$ and $b = \sin \frac{\theta}{2} \, e^{i\varphi}$,
and chain $C$ and qubit $B$ are in their ground states. Then we allow the system to 
evolve freely, so that $\ket{\Psi(t)} = e^{-iHt/\hbar}\, \ket{\Psi(0)}$ at time $t$.
The reliability of state transfer can be expressed in terms of the fidelity
\begin{align}
F_B(t) = \bra{\phi_T}\, \rho_B(t)\, \ket{\phi_T}\,,
\label{Eq:fidelity}
\end{align}
where $\rho_B(t) = {\rm Tr}_{A,C}\ket{\Psi(t)}\bra{\Psi(t)}$ is the reduced density 
matrix of qubit $B$ obtained by tracing out the degrees of freedom of $A$ and $C$ 
from $\ket{\Psi(t)}$, and $\ket{\phi_T} = a\ket{0}_B + b\ket{1}_B$ is the target state. 
We define ``state transfer" to occur at time $t_0$ when $F_B(t)$ attains its first maximum.
PST corresponds to the case $F_B=1$, where $t_0$ must be independent of the initial
state. For a uniformly coupled spin chain ($\lambda_\alpha = 1$), $t_0$ and the fidelity 
are both found to depend on the initial state. However,
we will show that PST can be attained with arbitrary accuracy in a spin bus geometry 
with $\lambda_\alpha < 1$, at a time scale $\sim 1/\lambda_\alpha^2$.

\paragraph{Quantum state transfer through odd-size chains.---}
An odd-size Heisenberg open chain has two-fold degenerate ground states, and behaves 
like a single spin-1/2 object at low energies~\cite{Friesen07,Oh10}. When qubits $A$ 
and $B$ are weakly coupled to the $i$-th and $j$-th spins of chain $C$, the chain acts 
as a ``central spin." The effective Hamiltonian \cite{Oh10,Shim10} can be computed to 
first order in the perturbation $\lambda_\alpha$ using the Schrieffer-Wolff 
transformation~\cite{Schrieffer}:
\begin{align}
H_{\rm eff} = J_{A,i}^*\,\bS_A\cdot\bS_{C} + J_{B,j}^*\,\bS_B\cdot\bS_{C}\,,
\label{Eff_Hamil_odd}
\end{align}
where $\bS_{C}$ is the central spin operator acting on the two-fold degenerate ground 
states of the chain, $\{\ket{0_C},\ket{1_C}\}$. The effective coupling to qubit $A$ is 
given by $J_{A,i}^{*} = {J_{A,i}}\,\bra{0_C}\sigma_{iz}\ket{0_C}\propto \lambda_A$, 
where $\bra{0_C}\sigma_{iz}\ket{0_C}$ represents the local magnetic moment at the $i$-th 
site of the chain. Due to the antiferromagnetic nature of the bare couplings, the 
effective couplings $J_{\alpha,i}^*$ alternate in sign as a function of the site position,
$i$. From now on, we assume qubits $A$ and $B$ are either both attached to even nodes or 
both attached to odd nodes of the chain, so that their effective couplings are both
ferromagnetic or both antiferromagnetic. If the qubits are attached to mixed nodes on 
the chain, we find that PST is not possible. For simplicity here, we take 
$J_{A,i}=J_{B,j}$, so that $\lambda_\alpha = \lambda$.

\begin{figure}[htbp]
\includegraphics[scale=1.0,angle=0]{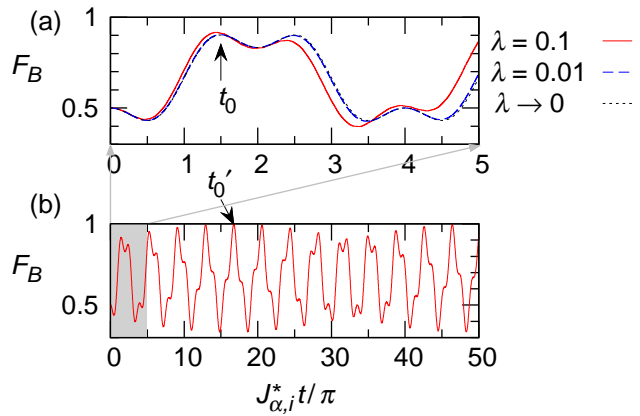}
\caption{(Color online).  PST fidelity $F_B(t)$ vs.\ time $t$ with an odd-size 
spin bus, for short (a) and long (b) time evolutions. The chain size here is $N=5$, 
and the initial state of qubit $A$ is 
$\ket{\psi_A} = \frac{1}{\sqrt{2}}\left(\ket{0_A} + \ket{1_A}\right)$. 
In panel (b), $\lambda = 0.1$.}
\label{Fig2}
\end{figure}

\begin{figure}[htbp]
\includegraphics[scale=1.0,angle=0]{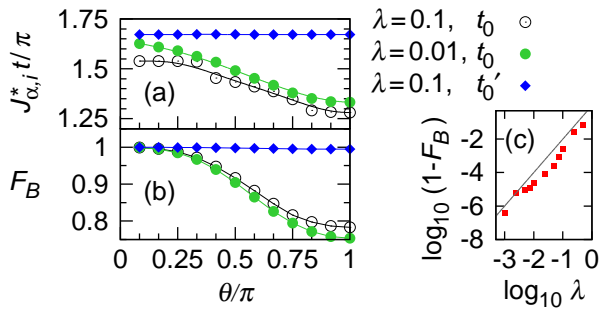}
\caption{(Color online). (a) Optimal times for state transfer, $t_0$ and $t_0'$ 
(scaled by $J_{\alpha,i}^*/\pi\sim\lambda$), as a function of the polar angle $\theta$ of 
the initial state, with an odd-size bus. (b) Maximum fidelity $F_B$ as a function of $\theta$. 
(c) Log-log plot of infidelity $[1-F_B(t_0')]$ vs.\ $\lambda$. The line represents $\lambda^2$.}
\label{Fig3}
\end{figure}

If switching of the bare couplings $J_{\alpha,i}$ is allowed {\it during} the evolution 
period, quantum state transfer from $A$ to $B$ can be achieved simply, through sequential 
swap operations involving the bus~\cite{Friesen07}. In the conventional PST protocol, 
however, the couplings must remain fixed.  Under this condition, it is impossible to 
transfer a quantum state from qubit $A$ to qubit $B$ perfectly, based on the evolution 
of the effective Hamiltonian~(\ref{Eff_Hamil_odd}). In this case, the fidelity can be 
computed exactly:
\begin{align}
F_B(t) = \frac{1+\cos\theta}{2}
      + \frac{\sin^2\theta}{4}\,f(t)
      + \frac{(1-\cos\theta)^2}{4}\,g(t) \,,
\label{Eq:Fid_odd}
\end{align}
where $f(t) =(5+ 4\cos6\tau + 3\cos 4\tau -12\cos 2\tau)/18$ and 
$g(t) = (7 + 2\cos 6\tau - 3\cos4\tau - 6\cos2\tau)/18$ with $\tau\equiv J^*_{\alpha,i}t/4$. 
$F_B(t)$ attains its initial maximum at $t_0\sim 1/J_{\alpha,i}^*\sim 1/\lambda_\alpha$, 
as shown in Fig.~\ref{Fig2}~(a). However, the maximum fidelity and the optimal time both 
depend on the state to be transferred, as shown in Fig.~\ref{Fig3}.  

While PST is not attainable using the effective Hamiltonian~(\ref{Eff_Hamil_odd}), this 
represents only the lowest order approximation to the full Hamiltonian (\ref{Hamil}) 
in the Schrieffer-Wolff transformation. When we compute the time evolution of the full 
Hamiltonian, we find that PST becomes possible. Figure~\ref{Fig2} shows the fidelity for 
short and long time periods for various $\lambda$. In the limit $\lambda \to 0$,
Eq.~(\ref{Eff_Hamil_odd}) becomes an excellent approximation and the fidelity follows 
Eq.~(\ref{Eq:Fid_odd}). On the other hand, as shown in Fig.~\ref{Fig2}~(b), 
for $\lambda = 0.1$, the higher order terms have a non-trivial effect on the time evolution, 
leading to a fidelity that approaches 1 at a new optimal time $t_0' \sim 1/\lambda^2$, 
which is about a factor $1/\lambda$ longer than $t_0$. More importantly, both the
maximum fidelity and the optimal time $t_0'$ are found to be independent of the initial 
state, as shown in Figs.~\ref{Fig3}(a) and (b), so that an unknown quantum state can be 
transferred perfectly. Figure~\ref{Fig3}(c) shows a log-log plot of $[1-F_B(t_0')]$ 
vs.\ $\lambda$. The infidelity has an approximate $\lambda^2$ dependence, suggesting 
that the second order terms play a key role in PST. In short, for an odd-size
spin bus geometry, $[1-F_B(t_0')] \sim \lambda^2$ and $t_0' \sim 1/\lambda^2$, so that 
PST can be achieved with arbitrarily high accuracy for smaller $\lambda$, albeit at 
the cost of a longer $t_0'$.

\paragraph{Quantum state transfer through even-size chains or rings.---}
Even numbers of spins with antiferromagnetic couplings are compensated, so the ground 
state of an even-size spin bus (chain or ring) is nondegenerate, and has no net 
magnetization. When qubits are weakly attached, an even-size chain or ring mediates 
an indirect exchange coupling between them. The effective Hamiltonian up to second order 
in the bare coupling strength is~\cite{Oh10}
\begin{align}
H_{\rm eff} = {e}_0\ketbra{0_C}{0_C} + J^{*}_{i,j}\, \bS_A \cdot \bS_B\,,
\label{Hamil_RKKY}
\end{align}
where the effective coupling is given by
\begin{align}
J^*_{i,j} = \frac{J_{A,i}\,J_{B,j}}{2}\, \sum_{n\ne 0} 
\frac{{\bra{0_C}\sigma_{i\mu}\ket{n_C}\bra{n_C}\sigma_{j\mu}\ket{0_C}}}{e_0 - e_n}\,.
\label{RKKY_coupling}
\end{align}
Here $e_n$ and $\ket{n_C}$ are the eigenvalues and eigenstates of an isolated ($\lambda = 0$) 
even-size spin bus, which we obtained numerically by solving the full spectrum of 
Hamiltonian ~(\ref{Hamil_chain}). Again we take $J_{A,i}=J_{B,j}$, so that 
$J_{i,j}^*\sim \lambda^2$. The time evolution of Hamiltonian~(\ref{Hamil_RKKY}) leads 
to a direct swap operation between qubits $A$ and $B$, and to PST. In this case, the
antiferromagnetic and ferromagnetic effective couplings are equally effectual for PST 
(in contrast with their ability to mediate ground state entanglement~\cite{Oh10}). 
The absolute value of $J^*_{i,j}$ determines the speed of the swap operation. The speed 
and fidelity of such quantum state transfer has previously been studied in a related 
but distinct geometry~\cite{Venuti07}.

In the limit of $\lambda \ll 1$, Eq.~(\ref{Hamil_RKKY}) forms an excellent approximation 
to the full Hamiltonian.  The fidelity of qubit $B$ is then given by
\begin{align}
F_B(t) = 1 - \frac{1}{2}\sin^2\theta \cos^2 \left(\tfrac{J^*_{i,j}\,t}{2}\right) \,.
\label{Eq:Fid_even}
\end{align}
In contrast with the case of an odd-size chain, PST now occurs at the first fidelity 
maximum, given by $J_{i,j}^*t_0=\pi$, and it is independent of the initial state. 
However, in this case, $t_0\sim 1/\lambda^2$, similar to $t_0'$ for an odd-size chain.

\begin{figure}[htbp]
\includegraphics[scale=1.0,angle=0]{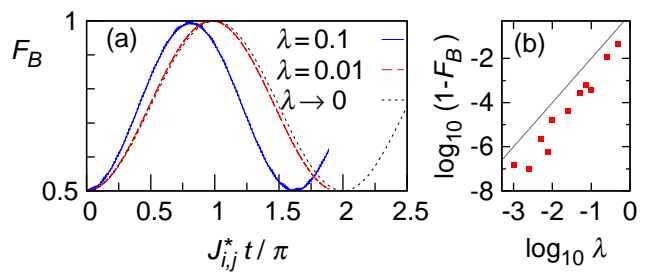}
\caption{(Color online). (a) Fidelity $F_B(t)$ as a function of time $t$ when qubits 
$A$ and $B$ are attached to the opposite ends of an even bus of size $N=4$. The initial
state of qubit $A$ is $|\psi_A \rangle = \frac{1}{\sqrt{2}}\left(\ket{0} + \ket{1}\right)$.
(b) Infidelity $[1 - F_B(t_0)]$ vs.\ $\lambda$. The line represents $\lambda^2$.}
\label{Fig:even1}
\end{figure}

Going beyond the lowest order effective Hamiltonian (\ref{Hamil_RKKY}), we can obtain 
exact numerical solutions using the full Hamiltonian, with the results shown in 
Fig.~\ref{Fig:even1}.  While panel (a) shows that $t_0\sim 1/\lambda^2$, panel (b) shows 
that PST occurs in the limit $\lambda\rightarrow 0$, with $[1-F_B(t_0)] \sim \lambda^2$.
 Both results have a direct correspondence with the odd-size bus geometry. We can 
therefore draw several conclusions. First, despite having very distinct ground states, 
even and odd-size buses both allow PST, over time scales of order $1/\lambda^2$. The 
similar behaviors seem to derive from the second order terms in the Schrieffer-Wolff 
transformation. Second, for larger $\lambda$, when higher order terms become important, 
PST is increasingly degraded. In this sense, it appears that the second order terms
are responsible for PST.

\begin{figure}[htbp]
\includegraphics[scale=1.0,angle=00]{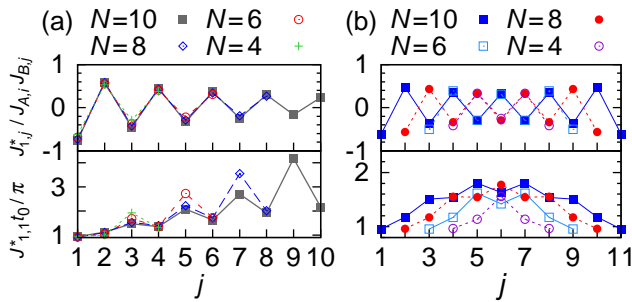}
\caption{(Color online). Effective couplings $J^*_{1,j}$, scaled by $J_{A,1}J_{B,j}$, 
and the corresponding optimal times $t_0$, scaled by $\pi/J^*_{1,1}$, as a function of 
position $j$ for even-size (a) chains and (b) rings of size $N$. Here, 
$J_{A,1} = J_{B,j}= 0.1$. In (b), plots with $N=4,6,8$ are shifted to the center, 
for comparison.}
\label{Fig:even_time}
\end{figure}

The time evolution of the even-size bus takes a simpler form than the odd-size bus, and 
the quantum interference is less pronounced. However, we now probe aspects of 
the evolution that are distinctly quantum in origin.  We study the dependence of PST on 
the separation distance between the attached qubits, for the case of an even-size bus. 
Here qubit $A$ is attached to the first node of the chain or ring, $i=1$, while qubit $B$ 
is attached to node $j = 1,2,\dots, N$ (see Fig.~\ref{Fig1}). For an open chain bus, 
the optimal time is found to increase as an oscillatory function of $j$, as shown in 
Fig.~\ref{Fig:even_time}(a). This leads to the striking observation that it can take 
less time to transfer a quantum state to a further qubit than a closer qubit.

Similar behavior is observed for an even-size ring bus, where an additional even-odd 
parity effect emerges, as shown in Fig.~\ref{Fig:even_time}~(b). This depends on the 
size of the chain: $N=2\times 2n$ or $2\times(2n+1)$. To see this, we first note that 
the ring has rotational symmetry. We might expect the transfer time to be maximized for 
a pair of antipodes, and this is true when $N=2\times 2n$.  However, when 
$N=2\times(2n+1)$, the PST time is maximized for the neighbors of the antipode. Note 
that the effective coupling between antipodes is antiferromagnetic if $N=2\times(2n+1)$, 
and ferromagnetic if $N=2\times 2n$. We conclude that quantum state transfer can be 
faster over longer distances than shorter distances. Such effects arise, in part, 
from quantum fluctuations of the bus eigenstates, and are distinctly quantum in origin.

\begin{figure}[t]
\includegraphics[scale=1.0,angle=00]{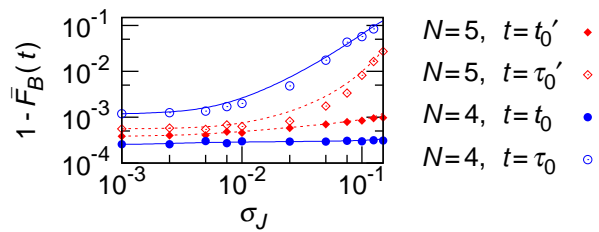}
\caption{(Color online). Average (over 100 random configurations with variance 
$\sigma_J$) PST infidelity $1-\bar{F}_B(t)$ with a 4- or 5-node bus as a function 
of exchange variation, represented by the variance $\sigma_J$ of a Gaussian 
distribution around the target $J_0$. Here $\tau_0$ ($\tau_0'$) is the PST time for 
the case of uniform exchange couplings, while $t_0$ ($t_0'$) refers to the PST time 
calibrated for non-uniform coupling strengths.}
\label{Fig:variations}
\end{figure}

So far we have focused on PST through spin buses with uniform Heisenberg exchange 
couplings. We have also studied how the fidelity of PST is influenced by random 
variations of the exchange couplings in the bus.  Consider the even-chain case as 
an example. The form of the effective second order exchange interaction, shown in 
Eqs.~(\ref{Hamil_RKKY}) and (\ref{RKKY_coupling}), does not change in the presence 
of small exchange variations in the bus, as long as the ground state of the bus 
remains nondegenerate. Thus PST should still be feasible, and its fidelity should
not change if the PST time $t_0$ (or $t_0'$) is calibrated to account for the variations 
in the effective qubit coupling strength $J_{i,j}^*$.  Without calibration, PST error 
should grow quadratically due to the sinusoidal time dependence of the PST fidelity 
(see Fig.~\ref{Fig:even_time}). Figure.~\ref{Fig:variations} clearly shows that after 
calibration, the PST fidelities for both even and odd size buses are indeed insensitive 
to variations in the exchange coupling. Without calibration, the PST error grows 
quadratically with timing error as expected. For variations $\delta J/J_0 < 0.01$, 
fidelity in both cases hardly increases compared to the perfect, uniform coupling 
result.

In conclusion, we have demonstrated perfect state transfer in a system where 
it was not expected: a strongly coupled Heisenberg chain or ring with uniform couplings. 
We have pinpointed the source of the effect in the second order terms in a Hamiltonian 
expansion, assuming perturbative couplings to the qubits. We have also shown that our 
PST protocol does not require precise engineering of the intra-bus exchange couplings, 
and that external qubits do not have to be coupled to the end nodes of the bus---they 
can be anywhere on an even bus, or on the same type of nodes (both even or both odd) 
on an odd bus. Moreover, we showed that for even-size chains or rings, the optimal time 
for PST is not linearly proportional to the geometric distance between sender and 
receiver, emphasizing the fundamental differences between quantum and classical signal
transmission. Overall, our results point to a truly robust coherent plug-in device 
for perfect quantum state transfer between remote spin qubits. The results presented 
here may be tested using existing technologies such as spin chains on a metal 
surface~\cite{Hirjibehedin06}, or with quantum dot~\cite{Petta05}, 
molecular~\cite{Timco09}, or NMR qubits~\cite{Cappellaro}.

\begin{acknowledgments}
This work was supported by the DARPA QuEST through AFOSR and NSA/LPS through ARO. 
L.A.W.\ was supported by a Ikerbasque Foundation Start-up, the Basque Government 
(grant IT472-10) and the Spanish MEC (Project No. FIS2009-12773-C02-02).
\end{acknowledgments}

\end{document}